\documentclass[11pt]{article}
\usepackage{chadstyle}  % Loads my formatting 
\usepackage{nameref}
\usepackage{hyperref}
\usepackage{graphicx}
\usepackage{fancyhdr}

\usepackage{amsmath,amssymb}

\renewcommand{\vec}[1]{\mathbf{#1}}
\begin{document}
\begin{singlespacing}
\begin{spacing}{0.9}
\begin{titlepage}

\title{A primer on model-guided exploration\\ of fitness landscapes for biological sequence design}
% \runningheads{A primer on model-guided exploration of fitness landscapes}{Sam Sinai and Eric Kelsic}

\author{\large {Sam Sinai\textsuperscript{1,2,3}} and 
\large{Eric D. Kelsic\textsuperscript{1, 2}}}
\date{}
\maketitle
\thispagestyle{empty}

%\clearpage
\vspace{-0.4in}

\begin{abstract}
Machine learning methods are increasingly employed to address challenges faced by biologists. One area that will greatly benefit from this cross-pollination is the problem of biological sequence design, which has massive potential for therapeutic applications. However, significant inefficiencies remain in communication between these fields which result in biologists finding the progress in machine learning inaccessible, and hinder machine learning scientists from contributing to impactful problems in bioengineering. Sequence design can be seen as a search process on a discrete, high-dimensional space, where each sequence is associated with a function. This sequence-to-function map is known as a “Fitness Landscape” \cite{wright1932roles,de2014empirical}. Designing a sequence with a particular function is hence a matter of “discovering” such a (often rare) sequence within this space \cite{povolotskaya2010sequence}. Today we can build predictive models with good interpolation ability due to impressive progress in the synthesis and testing of biological sequences in large numbers, which enables model training and validation.
However, it often remains a challenge to find useful sequences with the properties that we like using these models. In particular, in this primer we highlight that algorithms for experimental design, what we call “exploration strategies”, are a related, yet distinct problem from building good models of sequence-to-function maps. We review advances and insights from current literature -by no means a complete treatment- while highlighting desirable features of optimal model-guided exploration, and cover potential pitfalls drawn from our own experience. This primer can serve as a starting point for researchers from different domains that are interested in the problem of searching a sequence space with a model, but are perhaps unaware of approaches that originate outside their field.

\blfootnote{{Correspondence: sam.sinai@dynotx.com, eric.kelsic@dynotx.com 
\\
1:Dyno Therapeutics, Cambridge, MA 2: Wyss Institute For Biologically Inspired Engineering at Harvard Medical School, Boston, MA 3: Harvard University, Cambridge, MA}}

\end{abstract}

\end{titlepage}
\end{spacing}
\end{singlespacing}

% now start line numbers
%\linenumbers
%\pagebreak
\tableofcontents
\section{Preliminaries}
While we attempt to cover the advances that capture methods in approaching sequence design (or adjacent) problems, this is not a review and the field would always move faster than we can update this document.  We often find ourselves explaining different facets of the challenge to researchers from different fields, and hence we think it might be useful as a resource to share and lower the barrier for working on this problem, or perhaps help in making known approaches more accessible. As the purpose of this writing is to enable better algorithms,  we focus on methods (rather than results) when discussing approaches that were attempted to-date. 
\section{What is the problem?}

A ``fitness landscape" is defined as a map $\Phi: \vec{x} \rightarrow \vec{y}$ between biological sequences $\vec{X}=\{\vec{x}_1,\cdots,\vec{x}_n\}$ and their fitnesses $\vec{Y}=\{\vec{y}_1,\cdots,\vec{y}_n\}$\cite{wright1932roles}. Sequences $\vec{X}$ are each made up of alphabet (residues) $\Sigma=\{\sigma_1, \cdots, \sigma_k\}$ \footnote{$|\Sigma|=4$ for DNA and RNA nucleotides, and $|\Sigma|=20$ for standard amino-acids which are the building blocks of proteins}. Here, the term \emph{fitness} captures the biological desirability of a sequence. In nature, this can refer to the effect of multiple functionalities $\vec{y}= \{y_1, \cdots, y_m\}$ together on the organism's ability to survive or reproduce. For engineers, it might refer to other desired properties that don't necessarily align with biological fitness. The challenge for evolution or bioengineers is to find sequences $\vec{x}$ that exhibit some desired profile of functionalities $\vec{y}^*$ \cite{povolotskaya2010sequence}. The purpose of the primer is to introduce the reader to exploration algorithms $\mathcal{E}$ that query $\Phi$ as oracles to find sequences with the desired properties $\vec{y}^*$. Given that we do not know if sequences we are looking for actually exist, this may be an impossible task. However, assuming such profile exists, we aim to propose $\mathcal{E}$ that can find solutions within some acceptable distance of $\vec{y}^*$ with high probability.

In order to achieve this objective, it is often helpful to build a \emph{model} of the landscape, $\Phi'$ because accessing $\Phi$ directly is rather costly. Even evolutionary processes have an implicit model of fitness landscapes: They assume that the current members of the population are the most promising candidates to be  modified towards a better sequence, i.e. there is correlation between nearby sequences and their function. Engineers can build explicit models of the landscape $\Phi'$, which we will discuss in section 3.   

In its simplest form the challenge would be to find a single sequence that satisfies a single criterion. For example, the green fluorescent protein (GFP) is a 238 amino-acid protein that is a good experimental test-case for machine-guided modeling and optimisation \cite{sarkisyan2016local,biswas2018toward,brookes2019conditioning, otwinowski2018inferring, alley2019unified,biswas2020low}, and one may look to improve its fluorescence  ($y \in \mathbb{R}$), subject to the constraints that the protein still successfully folds.

\subsection{Why are biologists interested in building models of the landscape?}

While optimising a protein like GFP my not be very impactful \emph{per se}, as it is pretty good at what it does already, there are other natural proteins that perform the tasks we are interested in, but we would greatly benefit from optimising them. For instance, capsid engineering \cite{kelsic2019challenges}, a domain that we work on, stands to benefits massively from improved capsids that enable delivery of genes into specific tissues in order to treat currently untreatable genetic diseases. While the known natural variants of these capsid proteins show the desired traits, their current capabilities are far below the levels that make them appealing as therapeutics. 

Furthermore, as these landscapes are extremely large, we must understand a landscape's structure by sampling a very small portion of it. Evolutionary biologists are particularly interested in understanding landscape structure, because it can help make the process of evolution more predictable \cite{de2018utility,kryazhimskiy2014global,bank2016predictability}. Such understanding has theoretical  as well as practical implications. Practically, it can help us in estimating the probability of developing drug resistance for certain pathogens better, or even estimating the chances of a virus jumping from animal hosts to humans. More generally the structure of the landscape determines how hard it is for evolution to ``discover" novel functions.  These can help us progress towards understanding big scientific questions. For instance, we know that life on earth eventually discovered oxidative processes as a means of producing energy. Before that, Oxygen was very toxic to most living organisms, and the ``Great Oxidation Event" resulted in a mass extinction \cite{margulis1997microcosmos}. But it is by no means obvious how likely it was for evolution to learn how to use oxygen to make ATP (the cell's energy currency). Is the fitness landscape full of possible adaptations that allows for the use of oxygen? or is it that perhaps the ``fitness optimum" is atop a wide base where hill-climbing would get you there from many locations in the sequence space (however slowly)? Affirmation of either of these would suggest that these types of adaptations are probable. On the other hand, we often assume that a random sample of biological sequences is quite unlikely to contain sequences that exhibit a given function. In a famous experiment, Bartel and Szostak tried to provide some estimate of this challenge by screening a large pool of randomly synthesized RNA molecules, which were tested for their ability to ligate certain products\cite{Bartel93} . In their experiment, the frequency of detecting a functional sequence was roughly $10^{-14}$. 

To summarize, there is value in building better estimates of $\Phi$ for any landscape, first because it helps us optimize on the landscape, and second, because being able to estimate the landscape overall can help us understand the outcome of stochastic processes within it better. These are distinct problems, as in the second case, we would like the error of the estimator to be low on average, whereas in the optimization case, we simply care about finding high-performing variants. Depending on which objective we prioritize, or the type of model we desire, we might sample sequences within a landscape differently \cite{du2016good}.  

 We will call a sequence with a particular arrangement of alphabet a \emph{variant}. Often, in biology, the process of exploring a landscape involves trying many variants in parallel, which constitutes a \emph{population}. In natural (evolutionary) settings, it is often the case that multiple copies of the same variant exist in the population. In synthetic assays, it is more desirable to test unique variants (duplicates are included for calibration).  We also assume a fixed population size. In the approaches discussed here, we importantly assume that the fitness landscapes are \emph{static}, that is they don't change over time or as the frequency of different variants in the population change (those are called ``seascapes"). I.e. a population evolving over a landscape does not change the properties of the sequence-to-function map. Generally speaking, these are not true for natural organismal evolution, because the context changes the map, but hold for biological sequence design. The difference between $\mathcal{E}$ (or population) trying to optimize on static landscapes versus one that attempts to optimize on a dynamic one is like the difference between an indoor rock climber and a tennis player. In one case the challenge is somewhat fixed, while in the second it is opponent and tournament dependent.

\subsection{Why is this problem hard?}

The problem of optimizing sequences on fitness landscapes is hard because their map $\Phi$ often has complex, non-convex structure and landscapes are unimaginably large. Biological sequences often exhibit significant amount of \emph{epistasis}, which refers to the influence of genetic context on the effects of a change within a sequence. In the sequence design setting, if the effects of swapping one letter for another in a position (also termed \emph{locus}) in the sequence were independent of what happened elsewhere, we could feasibly measure the effects of changing a letter to another \footnote{Population geneticists call a particular pattern of letters within particular loci, an $allele$.}, for every position in the sequence (e.g. \cite{Ogden2019}) and then design the optimal sequence by picking the best letter at each position. Hill-climbing processes can optimize very easily in this setting as well \cite{wilf2010there}. While these types of landscapes may approximate the vicinity of a particular sequence well\cite{Ogden2019,bank2016predictability,otwinowski2018inferring,agarwala2019adaptive, de2018utility}, single-peaked landscapes are thought to be rare, and there is evidence to suggests that the global structure of landscapes can be quite heterogeneous \cite{bank2016predictability}.

Epistatic effects come in multiple forms, \emph{magnitude} epistasis refers to cases where one allele's presence changes the magnitude of the effect of another one (e.g. diminishing returns: if you have an apple, a second apple is less appealing to you). \emph{Sign} epistasis refers to situations where an allele's presence changes the sign of the effect of another allele (e.g. you are very sleepy, you drink coffee, you feel better, you now drink an espresso, you feel worse because you are anxious now, but no worse than you were before you had coffee). In the extreme cases, \emph{reciprocal sign} epistasis can lead to cases where two alleles that are good independently, end up being deleterious together (or vice versa, e.g. menthol and coke may taste fine by themselves, but having them together is extremely unpleasant). Due to epistatic effects, optimisation on biological landscapes involves dealing with many local optima. 

Secondarily, sequence spaces scale exponentially in sequence length $L$.  The total number of possible sequences of a certain length is $| \Sigma|^L$ where $L$ is the length of the sequence, and many sequences of different lengths may perform the same functions, often resulting in a huge space of possible candidates that exhibit similar functionalities.  Even to perturb a sequence with a few mutations, a combinatorial number of choices are possible, which means that even with a perfectly accurate model in hand, brute-force search is not feasible. Moreover, if the landscape is not dense in viable solutions, and exhibits sign epistasis, hill-climbing algorithms may be subject to exponential search times, even to find local optima \cite{kaznatcheev2019computational,chatterjee2014time}. It is natural to ask then, how and if evolution has solved this problem. There is evidence that evolutionary search is not an exceptionally great learning algorithm \footnote{For some static fitness landscapes Valiant \cite{valiant2009evolvability} showed that standard evolutionary processes cannot find optimal solutions when other learning algorithms do. A recent study by Kaznatcheev proposes that frequency-dependent dynamics (which we do not consider here) can overcome that barrier \cite{kaznatcheev2020evolution}. }, suggesting that the solution space cannot be too difficult (or otherwise we would not see evolution succeed). The most promising biological inventions that seem to remedy search times on such landscapes are recombination and the so-called  ``regeneration" processes where you reset your walks in the sequence space through gene duplication (ensuring you don't break your current best answer too much, and cap your distance from a potential solution) \cite{chatterjee2014time,mcdonald2016sex, sinai2018primordial}. 

\begin{figure}[ht!] 
\includegraphics[width=1\textwidth]{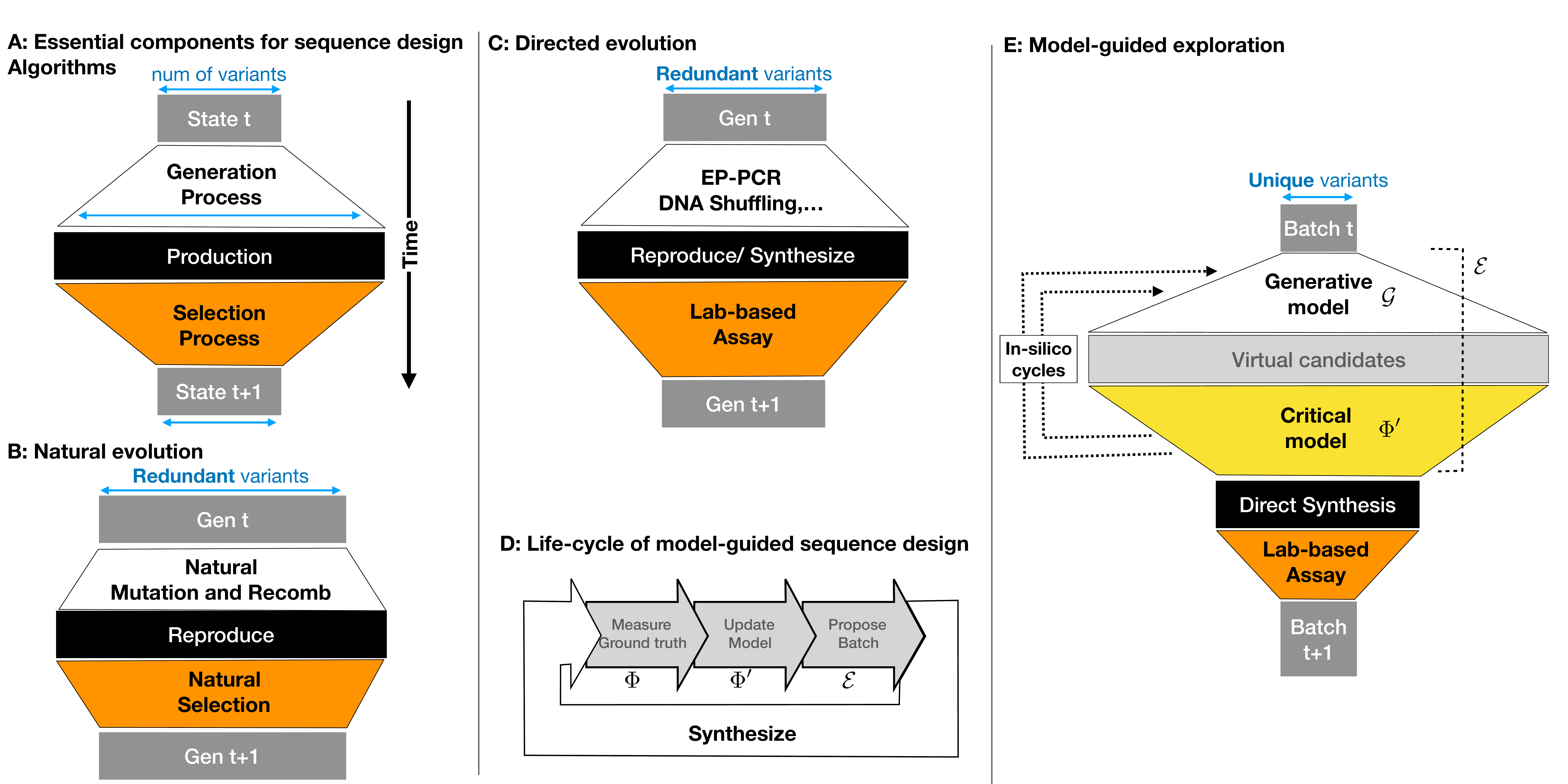}
\caption{ \textbf{A}: An exploration of the landscape has multiple stages. At each step, new candidates are proposed by a generative process, then they are produced and subsequently filtered by a selection process, which results in a new state for the population.  \textbf{B}: For natural evolution, mutation and recombination are the generative process, physical reproduction synthesizes the variants, and natural selection filters them (as the critique). \textbf{C}: In directed evolution, generative processes are often lab-based methods of diversity production (e.g. error-prone PCR or in-vitro culturing) and then lab based assays are used to critique the variants that were produced by those processes. \textbf{D}: There are four steps in the life-cycle of model-guided sequence design. Candidates are \textbf{synthesized} and then \textbf{measured} experimentally to gather information about the landscape $\Phi$.  Given this information, models of landscape $\Phi'$ can be trained, and used by $\mathcal{E}$ to explore the sequence space, by proposing new samples for synthesis. \textbf{E}: In model-guided exploration, candidates may go through the virtual generation and critique (using $\Phi'$) steps multiple times before synthesis. This increases the chance of producing samples that will succeed in lab-based assays, reducing experimental costs and accelerating discovery.} 
\end{figure}

\subsection{How can machine learning help in solving this problem?}

An exploration process on a fitness landscape can be described much like a Generative Adversarial Network (GAN) \cite{goodfellow2014generative}: (a) A generator that makes a set of proposal sequences (b) A critique that evaluates  the sequences produced by the generator (see Figure 1). Nature is the ultimate critique, but asking it questions is expensive. If we have a model that can propose functional sequences, then we can leave the job of the critique to nature. However, if our generator has a high rate of making poor choices (say it picks at random), then we would want to preface queries to nature by simulating the critique before committing to synthesizing new sequences. This involves building accurate models $\Phi'$ and using them as critiques. However, unlike GANs, error-propagation is not always possible throughout the entire chain, depending on our choice of generator and critique (you can't train nature's parameters). 

All algorithms (natural or otherwise) that explore fitness landscapes require both of these components. The trick for machine learning algorithms is to replace nature's generator to make better than random guesses or simulate its critique to screen sequences before they are made (of course in both cases, they need synthetic biologists to make them, i.e. do the expensive step of asking nature). 

In less metaphorical language, machine learning techniques that deal with exploration and exploitation trade-offs can help propose better sequences. Machine learning methods that help us model $\Phi$, either through black-box models or through better representations, can help us build better models of the landscape (oracles), thereby reducing the experimental costs. 

\section{How do we currently make models of fitness landscapes?}
 Fitness landscapes have been of interest to biologists for about a century\cite{wright1932roles}. There are two qualitatively distinct ways to approximate and study biological fitness landscapes: Empirical approaches based on experiments, and statistical approaches based on models which we will cover in detail. Empirical studies of fitness landscapes generally fall in two categories: First,  studies in which samples are taken from an evolving population under natural or directed selection. These studies provide the benefit of producing diverse and large datasets that can be generated at once with little cost \cite{good2017dynamics,venkataram2016development}. They also are relatively faithful models of evolutionary process. However, by definition, most of these studies explore the ``viable" regions of the landscape, leaving potentially interesting parts of the landscape untouched. It is hence difficult to know if a region of the landscape is non-viable or unexplored. Second, studies that sample variants using random or targeted mutagenesis \cite{barrera2016survey,de2014empirical,de2018utility,bank2016predictability,sarkisyan2016local, poelwijk2019learning,aguilar2017thousand}. In these studies one can synthesize a set of sequences and assay their performance in the laboratory.  Advances in sequencing technology has allowed approaches such as error-prone PCR  and DNA-reshuffling to generate sequences with random variation, where poor mutants can also be observed. Finally, direct synthesis approaches, the most recent development in the domain, allow for a large number of specifically designed edits to be synthesized and tested. The samples tested can be precisely selected, hence they are of high informational value.  However, mutational scanning is labor and resource intensive, and many orders-of-magnitude fewer samples are explored. 

It is transparent that the two approaches are synergistic and can be combined. The first step has been to use available experimental data to infer consistent parameters for each class of models (described below). Encouragingly, simple models such as Additive, low-order polynomial models, and global epistasis, have been well-suited to describe the local properties of experimentally probed landscapes\cite{otwinowski2018inferring,sailer2017detecting,hopf2017mutation,Ogden2019}. Due to multiple limitations in how well models can capture complex landscapes for which long-range regularity is not well-understood \cite{bank2016predictability,otwinowski2014inferring,sailer2018uninterpretable}, it is hard to predict the fitness of variants that are far from those already measured. A modern synthesis would likely involve mapping the insights gained from the empirical data back to refine the models that best describe these landscapes, hopefully in an interpretable, and generalizable manner. In this section, we will cover some classes of models that have provided useful insights over the years.

\subsection{Mathematical and statistical models}
For most of the time that the concept of fitness landscapes existed, measuring the effects of many mutations was hard. As a result, biologists have proposed multiple models of fitness landscapes that did not rely on empirical data, but aimed to emulate the properties of real landscapes. 

The most fundamental but simple models are those in which the effects of mutations are independent, hence the collective effects of all mutations can be computed through an additive function. These landscape set the benchmark to compare all other landscapes with because at the core, additive models capture the effects of a letter (allele) in a particular position (locus) when considered independent of each other.  Assuming we have a sequence $x := \{ \sigma_i \in \Sigma, \forall i  |  \sigma_1, \sigma_2, ...,\sigma_L\} $ of length $L$, the function (or fitness) $\Phi(x)$ can be defined as:

\begin{align} % requires amsmath; align* for no eq. number
\Phi(x)= \sum_{i=1}^L  \phi(\sigma_i) \omega_i 
\end{align}

If all weights $\omega_i=1$, and $\phi(\sigma_i)$ is the effect of a specific mutation at a specific location. This is the simple additive model\footnote{While generally used interchangeably, it's preferable to refer to the model as ``linear" when $\omega_i \neq 1$.}.  However as described above, biological landscapes often deviate from additive (or linear) models, and a variety of approaches are designed to capture this deviation based on different perspectives.

In the most basic form (sometimes used as the ``null'' model) epistasis is completely unstructured (random). This is termed House of Cards (HoC) epistasis. Landscape models can be built from a mixture of HoC and additive effect to incorporate measurement noise or otherwise incompressible fitness effects\footnote{In a interesting recent study, Agarwala and Fisher investigate the dynamics of evolutionary walks on such landscapes that are locally additive (and correlated) but become less correlated as the distance to each peak is increased \cite{agarwala2019adaptive}}. 

A generalization of additive models is known as ``Global epistasis" models where the additive baseline is transformed by a monotonic but non-linear function \cite{otwinowski2018inferring,sarkisyan2016local,tareen2020mave}\footnote{A neural network based implementation can be found here: \url{https://mavenn.readthedocs.io/en/latest/}}. Despite their simplicity, and since additive landscapes can be easily inferred, these landscapes are somewhat successful in describing empirical mutational scan data sets \cite{otwinowski2014inferring,de2014empirical,de2018utility, sailer2018uninterpretable, sailer2017detecting}, possibly because of the local regions that these scans tend to cover (there may be one dominant peak in the local neighborhood, as assumed in \cite{agarwala2019adaptive, bank2016predictability}), which could be separated by large ``valleys" \cite{pressman2019mapping}.  

\begin{align} % requires amsmath; align* for no eq. number
   \Phi(x)=\Gamma\Big(\sum_{i=1}^L  \phi(\sigma_i) \omega_i  \Big)+ \eta
\end{align}

Where $\Gamma$ is a non-linear function and $\eta$ represents HoC epistasis.

A different group of elaborations on additive models aim to capture higher-order epsitatic interactions by assigning (or inferring) weights associated with interactions. In the most common form, they are written as coefficients of a regression.  

\begin{align} % requires amsmath; align* for no eq. number
   \Phi(x)=\omega_0+\sum_{i=1}^L  \phi(\sigma_i) \omega_i  + \sum_{i=1}^{L-1} \sum_{j=i+1}^L   \phi(\sigma_i,\sigma_j)\omega_{ij}+ ...
\end{align}

Where $\omega_i$ are weights on individual loci (positions), $\omega_{ij}$ are weights on pairwise interactions and so on. Inspired by Ising and Potts models, these can be used as generative descriptions of landscapes with certain amount of interaction among positions, or with some tricks, the parameters $\omega$ can be approximated efficiently \cite{marks2011protein, ekeberg2013improved, hopf2017mutation, mann2014fitness}. However, as efficient inference requires that the parameters are limited to a certain degree of interactions, often only linear and pairwise interactions are considered. 

A more general approach introduced by Poelwijk and colleagues \cite{poelwijk2019learning}, emphasizes the role of background-averaging (observing mutations in multiple contexts, ideally all). This setup allows for the weights to be inferred across many degrees of interaction, decomposed through a Walsh-Hadamard transform (the discrete equivalent of a Fourier transform). The regression models can be imagined as the (local) Taylor expansion of the landscape centered at a particular reference, whereas Poelwijk et al's approach enables a global picture of the landscape akin to a Fourier decomposition \cite{poelwijk2019learning,weinberger1991fourier}. Provided such interactions are sparse across the sequence, they authors suggest that compressive sensing can be used to infer these weights with a partial measurement of the landscape, allowing prediction. 

Finally, a well-established set of landscape models is known as NK-landscapes \cite{kauffman1989nk}, in which $N$ is the length of the genome, and $K$ is the number of positions in that genome that affect the fitness contribution of a single locus. The $K$ parameter tunes the landscape for ``ruggedness", where $K=0$ denotes the additive model, and $K > 0$ describe landscapes with increasing number of local minima. While widely studied in the literature as they share qualitative features with empirical landscapes (and can be used to study \emph{classes} of landscapes \cite{kaznatcheev2019computational, kauffman1989nk}), they don't make good models of specific landscapes. Hence, they are more suitable for evaluating search processes across many different classes, however, it is unclear which ones would be biologically relevant, and even less clear if they are relevant for sequence design in a particular context. 

\subsection{Computational models}
These models are simulators of the behavior of particular set of biological sequences, and they primarily focus on biophysical structure. The simulators are built on thermodynamical principles or other domain knowledge. These landscapes show considerable epistasis and complexity akin to those observed in natural landscapes (at least for specific regions of the landscape) \cite{du2016good,otwinowski2014inferring}. The drawback for these models is that they work only for a subset of problems where the biophysical first-principles are indeed sufficiently explanatory. An example of such problems would be thermostability of proteins, although not all thermostability problems are well-captured by these methods.

Two of the most famous examples of these simulators are those that simulate RNA sequence secondary structures (e.g. the Vienna package \cite{lorenz2011viennarna}), and those that simulate protein stability and structure (e.g. Rosetta \cite{rohl2004protein}). While an algorithm like Rosetta can be an excellent tool for design of de-novo proteins \cite{Ng2019,Shen2018}, those that live in deep energetic valleys, it is less suitable at predicting the stability of natural sequences subject to perturbations, where the energy landscape is more subtle.  As their strength, the type of bias that these models rely on are (by definition) close to well-established natural laws. 

\subsection{Machine learning models}

A class of fitness landscape models, made available by the recent advances in deep mutational assays are fully data-dependent. These models, often black-box, are trained on data (both unsupervised and supervised), and are then used as an oracle for estimating fitness.  Unsupervised models make use of the vast amount of unlabeled evolutionary sequence data that is available publicly to infer the fitness landscape (the Potts-model inspired work above also fall in this category) \cite{riesselman2017deep, yang2019machine,rives2019biological}. The assumption fundamental to these models is that similar evolutionary sequences are roughly capturing the same function.  Supervised models depend on expensive assays to collect labels \cite{romero2013navigating,biswas2018toward, sarkisyan2016local,alipanahi2015predicting, wu2019machine, Ogden2019}, and often the assays are not designed for the models to optimally learn. In between these two limits there are semi-supervised approaches that try to take advantage of both sets of data \cite{alley2019unified, biswas2020low}. 

While it is possible to learn good models of the fitness landscape with these techniques, these black-box models are often not invertible and hence even with a perfect map of sequence to function, sequence design requires the ability to optimize. Additionally, the performance of these models are not uniform across the sequence space, and can vary drastically in different regions of the landscape. Therefore a desirable property of these models is to have them report their confidence in their estimates. These uncertainty estimates are inherent to models like Gaussian Processes \cite{romero2013navigating}, but can also be achieved with neural networks by using techniques such as ensembling or dropout\cite{lakshminarayanan2017simple}.

\section{What does a good solution look like?}
In this segment, we briefly cover the definitions and useful metrics by which sequence design algorithms may be evaluated. Often, sequence design is \textit{online}, meaning that decisions have to be made as we collect information, and hence in retrospect, better decisions might have been possible. In this context, an ``online algorithm" is an approach that provides certain guarantees for performance, whereas a `` heuristic algorithm" is one that provides no such guarantees. Unless a particular regularity in the structure of the fitness landscape can be exploited, the majority of algorithms that we will describe provide no performance guarantees.

Experimental biology is hard, and often to get any measurement of the ``ground-truth" function of a sequence takes significant time and effort.  It is possible to do these measurements in batches of a certain size. Hence, if one has a budget to test a total of $B$ sequences, and a maximum batch size of $b$, at each time step, the process gets to use the information gathered so far to propose sequences for the next batch. Practically, batch sizes can be of up to $10^6$ for DNA synthesis, and each round of experiments can take a few months, and hence algorithms that perform better with fewer batches are more practical.

At their core, these algorithms need to balance the exploration-exploitation trade-off, based on the number of experiments that are made available to them (the ``horizon"). Exploration can be seen as an information gathering act on the landscape, and exploitation is optimization of the desired outcomes based on that information.

\subsection{Outcomes}
Sequence design algorithms are considered useful if they can find a large number of distinct high performing sequences. We will break this heuristic down to two parts, optimality and diversity. These can be measured in multiple ways, and we discuss some simple metrics here.

For optimality, in most landscapes that these algorithms would be actually used for, critical information such as the best possible fitness $y^*$ or the set of all local maxima $\mathcal{M}$ is unknown (otherwise, why use an algorithm to find them).  Without loss of generality, we will assume that maximization is the objective, but it is noteworthy that while it is common to assume that the best sequence is the one with the highest value $\max(y)$, in reality the most desirable value of a trait is not necessarily the highest value. For instance, binding a particular target may be desirable, but binding it too strongly may be less desirable than binding it at a moderate level. 

If the target values were known for a landscape, an algorithm can be considered efficient in its own right if it could find sequences $\vec{x}_i \in \mathcal{S_\epsilon}$ such that $||\Phi(\vec{x}_i)-\vec{y}^*||<\epsilon$ with probability $p \geq 1-\gamma$ for some acceptable tolerance $\epsilon, \gamma \geq 0$ and some bounded budget $B(L,\Sigma)$. Ideally, we would like $|S_\epsilon| >> 1$. However, we often do not have this information about the landscape in hand. 

Without such information, one can assume the existence of a desired profile $\vec{y}^*$ and choose $\epsilon$ ad hoc (or perhaps anneal it) accordingly. Most approaches simply choose a single dimensional $y$ and assume $y^*= \infty$, which translates to optimizing for $\max(y)$ for some bounded budget and batch size. An expansion would be to consider the cardinality $|\mathcal{S}_{\tau}|$, where $\mathcal{S}_{\tau} = \{\vec{x}_{i} \mid \Phi(\vec{x}_{i}) > y_{\tau}\}$ and $y_{\tau} \in \mathbb{R}$ is some minimum desired value. Notably, different algorithms may be better or worse in different budget and batch size regimes. 

While finding a good solution is a primary goal, finding multiple solutions that are distinct is more desirable. This is because there are always reasons beyond the designer's control that can make a particular solution inadmissible. Having higher diversity hedges against this risk. Ideally, you don't get diversity by perturbing your best solution, but find very different sequences that are good at doing the same thing (find many maxima in multi-peaked fitness landscapes). 

Similar to the optimality case, if we knew $\mathcal{M}$ we could use the number of found maxima $|\mathcal{M}'_{y_{\tau}}|$ where $\mathcal{M}'_{y_{\tau}} \in \mathcal{M}$ is the maxima found above fitness $y_{\tau} $ by the algorithm as a measure of diversity. 

When we do not have access to the entire landscape, we can measure the diversity of sequences in $\mathcal{S}_{\tau}$. Unfortunately, summarizing the diversity of a set of sequences can be challenging when information about peaks is not available. High-dimensional clustering is has its own set of heuristic approaches, and we will not discuss them here. Simple metrics would include average pairwise edit distance, metric distance on embedded spaces, clustering, and site-specific entropy. None of these would uniquely describe the diversity of the set, and each are subject to their own drawbacks.

\subsection{Properties}
Apart from the outcomes of interest described above, there are also properties of the algorithm that would make it more practical or desirable. Depending on the type of problem, different properties could be prioritized. 

\subsubsection{Efficiency } We would like the algorithm to be efficient in two senses: (1) It should make fewer queries $q_{\Phi}$ to $\Phi$ (experimental samples) to achieve the same outcome and (2) We would like $q_{\Phi'}/q_{\Phi}$ to be small. This second property also relates to scalability as it determines the computational resources required for the design of each batch. 

\subsubsection{Scalability}
To be able to use our full experimental bandwidth our approach should be able to generate as many samples as our experimental batch of size $B$ can accept. Furthermore, algorithms should scale as $L$ grows. Note that it is possible to have algorithms that have low $q_{\Phi}$ and $q_{\Phi'}$, but are not scalable. For instance, they may be unable to produce enough diversity to fill a batch of $10^4$ sequences due to mode collapse. 

\subsubsection{Consistency} If our models get better, our performance should also improve, i.e. $\mathcal{E}$ should make use of the information that is available to it. 

\subsubsection{Independence} Ideally, $\mathcal{E}$ should be independent of the model noise and bias. The algorithms shouldn't assume a particular type of noise or bias and we should be able to change the model or underlying landscape without making the algorithms useless \cite{purohit2018improving}. I.e. if we switch the model class, we would still have a usable exploration algorithm.

\subsubsection{Robustness}
Closely related to independence, if the model is bad, (often because of misspecification \cite{otwinowski2014inferring})  the algorithm shouldn't fail completely. I.e. we would like the ``worst-case" performance to be reasonably robust\cite{purohit2018improving}.  

\subsubsection{Adaptivity} In the computational sense, adaptivity denotes how many processes you can run in parallel. In experimental language, it captures the penalty for sampling $N$ sequences at once (in one batch) as opposed to sampling them sequentially. The more adaptive a process is, the fewer samples need to taken serially, and hence, fewer batches are needed. 

\subsubsection{Reproducibility} We would like the algorithm to be reproducible. This means that it would be reasonably robust to hyper-parameters, easy to implement, and doesn't require extraordinary computational resources to perform well.

It is noteworthy that having an approach that excels in all of these criteria is often impossible. For instance, high adaptivity may come with the cost of lower independence overall (high adaptivity can be achieved by having an excellent model of the global landscape). 

\section{What approaches have been proposed to address this challenge?}
\subsection{Model-free approaches}
\subsubsection{Brute-Force search}

Due to the combinatorial explosion of the landscape size as a function of the sequence length, the brute-force approach is feasible only on tiny landscapes or well-defined subsets of the fitness landscape, (e.g. all single mutants \cite{Ogden2019}). The generative process can simply enumerate all possible candidates, and records nature's response. This is the only model-free algorithm with no stochastic element incorporated. All other well-known ``deterministic" search schemes (E.g. BFS, DFS, Beam Search, ...) employed in this context make some decisions stochastically due to the high-branching factor or have to limit the search to a very small and biased sliver of the sequence space.    

\subsubsection{Random synthesis}

On the other end of the spectrum, we can generate samples randomly. This is the equivalent of Bartel and Szostak experiments \cite{Bartel93}, where the generator is a random sequence generator, and the critique is nature. Given the very low success rate of the generator, this approach is impractical for optimization unless the problem is easy or the throughput is overwhelming.  

\subsubsection{Natural evolution}

Nature's way of exploring fitness landscapes is by evolution. Evolution's generator is random perturbations on the population that has already been successful and survived so far, so while locally (almost) random, it contains a lot of information that was accumulated during the course of evolutionary history (This information is used in many unsupervised algorithms). Once sequences are generated through this process, they are directly passed to nature to decide what survives. 
\\
There two famous models of evolutionary processes, the Wright-Fisher process and the Moran process\cite{nowak2006evolutionary}. In the Wright-Fisher process, at each step $N$ offspring are sampled by selecting a parent with probability proportional to their fitness, and replicating them (with some mutation rate $\mu$). The entire new generation is made up of the offspring produced in the previous step, hence the generations are non-overlapping.  

The Moran process \cite{nowak2006evolutionary} differs in that at each step only one member of the current generation is sampled with probability proportional to their fitness and replicated, and one is removed at random. Hence the generations (every $N$ replications) for the Moran process are overlapping. This results in a faster adaptation rate, as offspring in a given generation may descend from other high-fitness offspring within that same generation. 
\\
In both the Wright-Fisher and the Moran process, it is also possible to recombine with other offspring (or parent). This is in principle controlled by two parameters: (i) $r$ which denotes the probability of a cross-over between two sequences already chosen to be recombining (ii) $\rho$ which is the number of recombinations per offspring. It is common to set $\rho=1$, which enforces sex to be between only two parents (this is not the case with some viral reproduction and DNA shuffling approaches). For comparison to batched experiments, the Wright-Fisher process appears a more suitable benchmark, however we expect natural populations to adapt at a rate closer to that of the Moran process (which is faster). 

However, a more subtle aspect of fitness landscapes is the resolution at which two different traits would be considered different enough from evolution's perspective. Assuming at trait value $y$, we define the probability of sampling a mutant $i$ for reproduction as $p_i=\frac{e^{\beta y_i}}{\sum_j e^{\beta y_j}}$ where $\beta$ is the intensity of selection. Hence the relative fitness can be written as $e^{\beta (y_i-y_j)}$, and the minimum $\Delta_{ij} y = y_i-y_j$ for which selection overtakes genetic drift (effects from random sampling, independent of the phenotype) as the primary force is approximately $\frac{1}{\beta}\log(1+1/N)$. In other words, having a larger population (or stronger selection intensity) results in a landscape with higher resolution where a smaller difference in the trait value would be effectively ``seen" by evolution.
\\
Hence, evolutionary processes like the Wright-Fisher can be simulated with four parameters: population size $N$, mutation rate $\mu$ (which we define as the probability of an edit to a base or amino-acid), recombination rate $r$ (implicitly $\rho=1$), and selection strength $\beta$. Evolutionary processes are often simulated in the limit where selection is strong and mutations occur infrequently (known as the strong selection, weak mutation limit SSWM), such that the population is monomorphic and updates one variant at a time (a variant appears and is selected to fixation, rather than having many competing variants in the population at once). This simplifies theoretical analysis significantly. However, this setting is unsuitable for measures like batch efficiency as you need many generations to achieve a high fitness, but is relatively sample efficient if you only care about unique samples tested. 

Natural populations face the prospect of extinction. As a result, populations cannot tolerate very high mutation rates. For example small populations can quickly run out of good variants if the mutation rate is too high, resulting in further shrinkage of the population and exacerbation of the problem\cite{gabriel1993muller}. Even for large populations (e.g. viral populations), natural organisms show mutation rates below what is known as the \emph{error threshold} defined as $\mu < \frac{1}{L}$, where $L$ is the size of the genome \cite{eigen1971selforganization}.  But in our fixed population size models with no minimum fitness criteria, something will eventually survive, and hence it is noteworthy that optimal mutation rates in these conditions are possibly above the error threshold.  Aside from these limitations, evolutionary processes in nature often benefit from achieving extremely large $N$, something that is far more limited in all the following strategies\footnote{However, in nature, the census population size is different (larger) than the equivalent (effective) population size $N$ that we use in the WF process, due to factors like non-random mating.}.

\subsubsection{Directed (in lab) evolution}

Classic directed evolution techniques take the principles of evolutionary search described above and apply it to problems of interest in lab. However, they also eliminate some of the constraints on natural evolution. They share the aspect that the generator is a set of random perturbations on an already functional population. However, in lab, we can increase the recombination frequency, i.e. $\rho >> 1$ and $\mu >>1/L$ to generate farther diversity, as we get second attempts if a population of mutants is completely dead.  On the critique side, they often bias and enhance the selective force $\beta$ toward a particular phenotype of interest. This is a double-edged sword, on one hand this allows the power of evolution to be focused on a trait of interest, on the other hand it may result in deterioration of secondary traits that are also important and multi-objective directed evolution is challenging (see \cite{nourmohammad2019optimal} for a modern framework towards this).

While these family of approaches have been extremely successful (and Francis Arnold won a Nobel prize for pioneering them \cite{arnold1998design}), it's been obvious for a while that at least on the choice of the what population to use as parents, augmenting them with a model (critique) can help with efficacy of the process \cite{romero2013navigating, fox2005directed, wu2019machine}. 

\subsection{Model-guided approaches}

While the baselines introduced above are ones that have been the state of the art for many decades, with the advent of DNA synthesis technologies and rapid increase in the accessibility of machine learning, there is a lot of activity in the field of machine learning for protein engineering\cite{yang2019machine,wu2019machine}. The ambition is to surpass evolutionary search in how quickly we can find good solutions \cite{kelsic2019challenges}, and further decrease the time and effort required to do so (Note that all ``model-free" approaches can be applied to optimise on $\Phi'$ instead of $\Phi$).  

\subsubsection{In-silico evolutionary algorithms}

Among the first studies that incorporated the entirety of the generate-model-explore framework was a series of papers by R. J. Fox and collegues \cite{fox2003optimizing, fox2005directed, fox2007improving} which first developed the concepts on NK-landscapes and then tried them on empirical ones. In these papers, the authors used in vitro and in silico \textbf{genetic algorithms} (random mutation and recombination) as their baseline generator in combination with Partial Least Squares (PLS) regression. In subsequent rounds used the weights of their model to inform the synthesis of mutants, hence, the model was employed both on the generator and critique side. Evolutionary algorithms are often a great choice to use for optimising on the landscape estimate. They can also be used in tandem with simple models (e.g. by structure \cite{bedbrook2017structure}). The field of genetic and evolutionary algorithms is massive, and applied in many contexts \cite{salimans2017evolution,deb2001multi,back1996evolutionary}. We encourage the reader to explore the resources on quality-diversity algorithms linked in section 7. However, most of these efforts are done outside sequence design and have not been re-applied to biology itself, partially because they historically preceded the technical ability for direct synthesis.  We have revisited this gap recently, and shown that simple and scalable evolutionary algorithms are competitive with the more modern approaches in terms of performance, robustness, and consistency \cite{sinai2020adalead}. These results suggest that evolutionary algorithms are still a benchmark to consider for model-guided sequence design.

\subsubsection{Adaptations of classical explore-exploit algorithms}

Sequential learning algorithms were developed in context where decisions have to be made based on partial observation of the data, and where choices may further inform the algorithm about the problem. In this online setting, the agent (explorer) needs to navigate the trade-off between exploiting the current best known solutions, and exploring new solutions with the hope of improving on the best solution.

Among the most applied approaches adapted from this class of algorithms is the framework of Bayesian Optimisation (BO). BO algorithms are designed to optimize (possibly non-differentiable) black-box functions that are expensive to query (which is the setting we are in). Importantly, these algorithms make use of the uncertainty of model estimates to negotiate exploration vs. exploitation. As the algorithm starts out with no knowledge about the space we are trying to optimize on, a prior over the set of possible objective functions is assumed (hence Bayesian). The goal is then to update the prior belief with measurements to obtain a posterior distribution over the set of functions. The function that approximates the posterior distribution of $y$ is known as the surrogate model (a more informative $\Phi'$). Most BO approaches employ Gaussian Processes as their surrogate (see this tutorial: \cite{brochu2010tutorial}). 

An important aspect of optimization in this setting is to decide how to use uncertainty to prioritize measurements. In BO, this is termed as the ``acquisition function". Some widely used acquisition functions are  (i) Expected Improvement (EI) which picks the sample that has the highest expected improvement over the current best sample and (ii) Upper Confidence Bound (UCB) where the expected reward is augmented with an additional ``optimism" term, for instance, by adding $\mu_r+ k \sigma_r$ where $\mu_r$ is the mean posterior,  and $\sigma_r$ is the standard deviation of the posterior, and $k$ is some scalar. Samples in UCB are then chosen as the one with the highest value of $\mu_r+k \sigma+r$.

In a pioneering study, Romero and colleagues \cite{romero2013navigating} demonstrate the use of BO for protein engineering. Many productive efforts have followed since \cite{bedbrook2017machine,gonzalez2015bayesian,yang2019batched} and are well covered in this review \cite{yang2019machine}.  While BO is a principled approach for optimization of black-box functions, it scales poorly in high-dimensional and high-throughput domains, such as those frequently encountered in sequence design. Specifically how to define the domain of optimization, adapt to batch setting,  represent sequence space, and choose good hyper-parameters can be challenging and differ significantly between experiments. Successful applications of BO to sequence design often require some domain art.  

In the recent decade, another domain of study with applications to black-box optimization has flourished: Reinforcement Learning (RL). RL algorithms learn to perform tasks by experience, hence their success is often dependent on whether interactions with the environment are cheap or if there is a good simulator of the environment in which they can practice (e.g. chess). The ``agent" that is guided by the algorithm interacts with the environment (or simulator), and observes rewards for taking different actions, over time, the agent learns to take actions with better reward. In our setting however, good simulators often don't exists, and sampling the environment directly is very expensive. 

One exception is that of RNA secondary structure, for which good simulators exist (e.g. \cite{lorenz2011viennarna}). Eastman et al \cite{eastman2018solving} take advantage of this to train reinforcement learning agents that are able to fold RNA sequences into particular secondary structures that are challenging to achieve. Most standard approaches to this problem include stochastic search (but generating random perturbations of current candidates), which results in inefficient sampling and reduced performance. Eastman et al, instead use graph-convolutional neural network to propose samples based on the networks knowledge of the simulated RNA-folding data, and train it with a reinforcement learning algorithm known as Asynchronous Advantage Actor-Critic algorithm (A3C). 

Another promising approach is to build locally accurate simulators and use them to train an RL agent. Angemueller et al \cite{Angermueller2020Model-based} train a policy network (a network that decides what mutations to make), by simulating the fitness landscapes through an ensemble of models. The models are trained on the measured data so far, and those that achieve high $R^2$ in cross-validation are selected as ``simulators". Agents are trained within this simulator up to the certain distance (determined by uncertainty in model estimates) from where data exists. Additionally, to increase the diversity of proposed sequences, they add a penalty for proposing samples that close to previously proposed samples (the closer they are, the higher the penalty).  

The drawback of reinforcement learning approaches is the high computational (and domain expertise) cost of using these algorithms. In particular, PPO and TRPO algorithms are known to be highly implementation-sensitive \cite{engstrom2019implementation}.

\subsubsection{Exploration on compressed representations of sequences}

Biological sequences show a significant degree of regularities (e.g. motifs appear in sequences, which result in structural or functional properties). Hence, in principle the space of sequences that we are interested in can be compressed into a different representation, ideally continuous (an ``embedding"), where we can use gradient-based optimisation techniques, or at least reduce the dimensionality of the data. Due to intense interest, these embedding methods are rapidly evolving, as they are also suitable for ``semi-supervised" settings, where the embedding is learned using unlabeled natural sequences, and subsequently supervised models are trained on the smaller dataset (now embedded), with better performance.  

Schemes to embed biological sequences fall into several general categories, notably VAEs \cite{gomez2018automatic, sinai2017variational, riesselman2017deep,ding2019deciphering,greener2018design} and invertible generative models (e.g. Flows or realNVPs) \cite{noe2019boltzmann}, as well as a plethora of increasingly promising models that have been adapted from natural language processing \cite{yang2018learned,alley2019unified, rives2019biological,rao2019evaluating,madani2020progen}.  However, the exploration strategies implemented on these learned embeddings follow familiar Monte-Carlo, hill-climbing, or Bayesian Optimisation schemes. I.e. these embeddings simplify the exploration problem itself, rather than employing a complex optimisation on the original sequence space.

\subsubsection{Regularized generative methods}

The use of generative models to propose sequences with better properties is a natural route towards achieving the objective of producing diverse sequences that we did not observe within the training set. We highlight some of these approaches below.

Brookes and Listgarten \cite{brookes2018design} approach the exploration problem by carefully pairing a generative model with a regressor (oracle) that guides the generator towards it's own optima. Their first algorithm, titled Design by Adaptive Sampling (DbAS), assumes access to a static $\Phi'$. Their algorithm works by training a generative model $G_{\theta}$ on a set of sequences $x_0$, and generating a set of proposal sequences $\hat{x} \sim G_{\theta}$. They then use $\Phi'$ to filter $\hat{x}$ for their high-performing sequences, retrain $G_{\theta}$ and redraw samples and iterate until convergence. This scheme is identical to the cross-entropy method with a VAE as the generative model (although they could use another generative model), an important optimization scheme \cite{CEM}. Notably, the oracle is not updated during the process. I.e. out-of-the-box their process is described with two round of experiments in mind (a training set to make an oracle, and a resulting set proposed by the generative model), where they maximize the potential gains from their oracle, given what it already knows. While it is trivial to repeat the process for multiple rounds, the process can be improved by incorporating information about how many rounds it will be used for. 
\\
In follow-up work \cite{brookes2019conditioning}, Brookes et al, aim to improve the robustness of DbAS by introducing CbAS. This is meant to address the pitfall in which $\Phi'$ is biased, and gives poor estimates outside its training domain. The authors enforce a soft pessimism penalty for samples that are very distinct from those that the oracle could have possibly learned from. Specifically, they modify the DbAS paradigm such that as the generator updates its parameters $\theta_0 \rightarrow{\theta_t} $ while training on samples in the tail of the distribution, it discounts the weight of the samples $x_i$ by $\frac{\mathbb{P}(x_i|G;\theta_0)}{\mathbb{P}(x_i|G;\theta_t)}$.  In other words, if the generative model that was trained on the original data was more enthusiastic about a sample than the one that has updated according to the oracle's recommendations, that data point is up-weighted in the next training round (and \emph{vice versa}). 
\\
Feedback-GANs (FBGAN) \cite{gupta2018feedback} approach sequence design using a GAN, which consisted of a generator $G$, and a discriminator $D$. In a standard GAN the generator's objective is to use a latent code $z$ (often white noise) to produce samples $x$ that are close to training data. The discriminator's objective is to discern if the $x$ was generated by $G$ or came from the ground truth distribution $R$ (i.e. $D(x_i)=\mathbb{P}(x_i \in R))$. By jointly training $G,D$ one can generate samples similar to those in the training set. Note however, that the discriminator does not necessarily provide an estimate of $\vec{y}$. To allow for optimizing $\vec{y}$, the authors pair the GAN with an oracle $\Phi'$. At each epoch, multiple samples generated by $G$ are passed through the oracle to acquire their label. Sequences with $\vec{y}>T$ (e.g. top quantile),  are then presented to the discriminator for training. This ensures that over time, both $D,G$ are biased towards proposing high-performing samples. 

Killoran et al \cite{killoran2017generating} pursue the same objective by regularizing an ``inverted model", $\Phi'^{-1}(\vec{y})=x$ through a generative adversarial network $G$. Activity maximization is a process where the input $x$ is perturbed in small steps such that it improves the objective $\vec{y}$, i.e. $x^{(t+1)}=x^{(t)}+\alpha \nabla_x \vec{y}$ ($\alpha$ is the step size). For sequences, due to their non-continuous nature, this involves continuous relaxation schemes of the input through casting them as probabilities: going from one-hot representations of sequences to position weight matrices (PWM). Activity maximization processes come with potential drawbacks however: (i) They can produce non-realistic sequences that do not look like those in the training data, (ii) they can be computationally expensive. The authors address the first problem by pairing the model with a generator (in their case a Wasserstein GAN) that is trained to produce samples similar to those that have the desired properties in the training set. The WGAN accepts a latent code $z_i$ to generate sequence $x_i \sim G(z_i)$, and then pass them to $\Phi'(x_i) = \vec{y}$. The optimization recipe then becomes $z^{(t+1)} = z^{(t)}+ \alpha \nabla_z \vec{y}$ where $$\nabla_z \vec{y}= \sum_{d} \frac{\partial \vec{y}}{\partial x_d}\frac{\partial x_d}{\partial z}$$ 
where $d$ represents the dimensions of $x$. 

Deep Exploration Networks (DENs) \cite{linder2020generative}, take a similar architecture as Killoran et al. but focus on optimizing the generator $G$ only. They assume access to an oracle $\Phi'$ (pre-trained) that they do not optimize during training (i.e. the oracle training time is the offline overhead, which saves a lot of time compared to computing AM online). The main innovative aspect of this work is to force the generator to compete with itself to maintain diversity: a known pathology of GANs is to ignore the latent code and undergo ``mode-collapse"  (where the generator simply produces one type of sequence). To achieve this, the authors provide distinct latent codes to $G$, and penalize outputs $x_1 \sim G(z_1)$, $x_2 \sim G(z_2)$  that are similar. The cost function for this approach is defined as :
$$C = C_{objective}\Big(\Phi'(G(z_1))\Big)+C_{diversity} \Big(G(z_1),G(z_2)\Big)$$
which they seek to minimize by training $G$. This approach circumvents some of the drawbacks of AM, notably the computational cost of producing sequences. A simpler architecture with fast convergence has been proposed in follow up work \cite{linder2020fast}. 

\subsection{Meta-algorithms}
Of course, one is not restricted to use only one of the methods above to propose sequences. Angermueller et al. \cite{angermueller2020population} propose a population-based method (P3BO) that ensembles exploration strategies together. The idea is very simple: at first many exploration strategies are given equal shares of the budget for each batch. Subsequently, given the performance of the sequences each proposed, the budget allocated to each method is updated, rewarding algorithms with better proposals. They show that this ensembling generally outperforms the baselines that only include single methods both in terms of diversity and optimality. 

Another innovation that can be thought of as a meta-algorithm, applicable to many generative exploration strategies, is termed \emph{auto-focusing} \cite{fannjiang2020autofocused}.  When a model is trained based on some collected data, and new samples are desired for the next batch that somehow optimize the properties of the input. Often, as discussed in this primer, the trained model is misspecified and starts to become inaccurate as the inputs deviate from those present in the training data. Search algorithms discussed above that query such models define a ``trust region" (which can be soft or hard) and only accept designed inputs if they fall within that region. Auto-focusing is a recipe to instead retrain the predictive model using importance weighting, coupled with updates to the parameters of the search model (i.e. the generative explorer) in lock-step, to ``focus" the model on the region of the search space that the explorer is visiting (i.e. improve the model's accuracy in that region), thereby improving the quality of the model (reducing its gap from the ground truth) in the region that the explorer is sampling.

\subsection{Potential pitfalls}
In this section, we list some considerations that are occasionally ignored in studies that have significance for evaluation and designing algorithms that can be used in practice. 

\subsubsection{Optimizing with pathological oracles}
Ground truth oracles that are only proxies of ground truth (e.g. machine learning model of the landscape), can behave pathologically outside their training domain \cite{kumar2019model,brookes2019conditioning, Angermueller2020Model-based}. Algorithms that explore spaces built on these models can overfit in such landscapes: the biases introduced by these oracles can make an algorithm appear good in practice, when it is in fact a good optimizer for the model. 

It is hence advisable to use ground-truth oracles that with consistent behavior across the entire exploration domain\cite{sinai2020adalead}. When not possible, a compromise would be to ensure that the algorithm's internal model is not from the same class as that of the surrogate oracle \cite{brochu2010tutorial}. 

\subsubsection{Allowing duplicate sequences for synthesis based approaches}
When synthesis is the way in which sequences are generated, populations of sequences shouldn't contain duplicates, either within or between generations (except for experimental controls). Experimentalists don't need to re-measure sequences once they have a good measurement. This could artificially inflate diversity, and mean fitness measures.  Natural and model-free directed evolution benchmarks are an exception. 

\subsubsection{Only testing on landscapes with gradients everywhere}
A challenge that sequence design algorithms face in practical settings is that generating samples with aggressive changes may result in a batch of completely non-functional sequences. This would be costly and doesn't add much information to the models in order to improve the guesses for the next round.  However, when test landscapes contain gradients everywhere, mistakes are not as costly, and algorithms can always re-initiate at a random place and optimize successfully. Testing on landscapes of this second kind only, can give results that are not generalizable to more realistic landscapes where vast areas of the sequence space contain no gradients (we term this as ``swampland", but they've also mentioned as ``holey" \cite{romero2013navigating}). 

\subsubsection{Initializing optimization at random locations}
This is related to the previous pitfall. Many realistic landscapes are swamplands, where random samples of sequences will not exhibit any function at all \cite{Bartel93}. In such landscapes it is obvious to the designer that they need to start at sequences that already show some desired activity. However, this means that those sequences may already be at or near a peak. Algorithms that are very good hill climbers may not be great at escaping local peaks. Hence even when landscapes have gradient everywhere, it is advisable to test the algorithm on both high-performing starting samples as well as random (often poor) samples.   

\subsubsection{Miscalibration of selection pressure or population size when comparing with evolution}
As discussed in section 5.1.3, the selective pressure and population size are highly important in the efficiency of evolutionary exploration. In particular, for any given population size, there fitness landscape is coarse-grained with respect to the minimum difference in fitness $\Delta y$ under which selection overtakes drift. In other words, evolution cannot ``see" the difference between mutants that have $\Delta y <<\frac{1}{\beta} \log (1+1/N)$. For this reason, one measure of efficiency or optimally, can be to state how fast a population reaches the same fitness that an evolving population with size $N$ and selection intensity $\beta$ would.  

\section{Conclusions}
Sequence design with the aid of machine learning is a booming field that spans multiple disciplines. The topics covered in this primer are meant to educate and help initiate research from experts and beginners across these fields, and ease their introduction to this topic. This primer does not paint a full picture, and in particular, is light on its treatment of computational protein design. Additionally, given the rapid expansion of the field, we did not attempt to cover all recent progress in this domain. However, we hope this resource removes some of barriers for those entering the field. Fortunately, there is a lot still to be done, and we look forward to the upcoming progress.
\clearpage
\section{Further resources}

\begin{itemize}
    \item A list of quality-diversity algorithms:
    
    \url{https://quality-diversity.github.io/papers}
    \item A regularly updated repository of papers related to Model-guided protein design:
    
    \url{https://github.com/yangkky/Machine-learning-for-proteins}.
    \item A sandbox for evaluating sequence design algorithms:
    
    \url{https://github.com/samsinai/FLEXS}.
\end{itemize}

\section{Acknowledgements}

We thank Jeff Gerold, Lauren Wheelock, Carl Veller, Gleb Kuznetsov, Surge Biswas, Martin Nowak, George Church, and members of Dyno Therapeutics for helpful discussions.

\clearpage

%\nolinenumbers

%This is where your bibliography is generated. Make sure that your .bib file is actually called library.bib

\end{document}